\documentclass[aps,prc,floatfix,superscriptaddress,preprint,showpacs]{revtex4}
\usepackage{times}
\usepackage{dcolumn}                 
\usepackage{graphicx}
\usepackage{epsf}
\usepackage{amsmath}
\usepackage{bm}
\newcolumntype{.}{D{x}{}{-1}}

\newcolumntype{w}[1]{D{.}{.}{#1}}

\begin{document}
\preprint{Version 1.0}

\title{Radiative orbital electron capture by the atomic nucleus}

\author{K. Pachucki}
\affiliation{
Institute of Theoretical Physics, Warsaw University,
Ho\.{z}a 69, 00-681 Warsaw, Poland}

\author{U.D. Jentschura}
\affiliation{Max-Planck-Institut f\"ur Kernphysik, Saupfercheckweg 1, 69117 Heidelberg}

\author{M. Pf\"utzner}
\affiliation{
Institute of Experimental Physics, Warsaw University,
Ho\.{z}a 69, 00-681 Warsaw, Poland}

\begin{abstract}
The rate for the photon emission accompanying orbital 1S electron capture by the
atomic nucleus is recalculated. While a photon can be emitted by the electron or by the nucleus,
the use of the length gauge significantly suppresses the nuclear contribution.
Our calculations resolve the long standing discrepancy of theoretical predictions
with experimental data for $\Delta J=2$ forbidden transitions.
We illustrate the results by comparison with the data established experimentally
for the first forbidden unique decays of $^{41}$Ca and $^{204}$Tl.
\end{abstract}

\maketitle

\section{Introduction}
Orbital electron capture by the atomic nucleus (EC) is one of nuclear decay
modes governed by weak interactions, a common and well known type of
radioactivity \cite{bambynek}. In this process the released energy $Q$ (equal
to the transition energy $Q_{EC}$ minus the binding energy of the captured
electron in the daughter atom) is shared between an emitted electron neutrino
and the final atom. With a small probability, of the order of $10^{-4}$ with
respect to the normal EC decay, a photon can be emitted together with the
neutrino. In such radiative electron capture decay (REC) the energy is
shared statistically between three bodies, thus the energy spectrum of
these photons is continuous.

The theoretical description of REC was initiated by
Morrison and Schiff \cite{mor_schiff} who neglected the Coulomb field of
the nucleus and took into account just 1S electrons and only in non-relativistic
approximation.
An advanced and more accurate theory of radiative electron capture was
developed by Glauber and Martin in Refs. \cite{glauber, martin}.
They included exactly the Coulomb field in the propagation of the electron,
took into account relativistic effects as well as screening and considered
captures from higher shells. However, they limited themselves to the
allowed decays where the nuclear spin changes by $\Delta J=0,1$ with no parity
change. Predictions of this model
were tested in a large number of experiments, and satisfactory agreement
was found \cite{bambynek}.

A more general theory of REC, extended to any order of forbiddenness (i.e.
for arbitrary change of nuclear angular momentum and parity), was developed by
Zon and Rapoport in Ref. \cite{zrap} and Zon in Ref. \cite{zon}. For verification of their
results, first forbidden unique transitions ($\Delta J=2$, $\pi_i \, \pi_f =-1$)
are of special importance because of a cancellation of the nuclear matrix
elements in the ratio of radiative to the nonradiative capture rate.
Measurements of radiation accompanying the 1S EC decay in the case of
$^{41}$Ca \cite{41Ca}, which belongs to this category, revealed a serious disagreement  with
results of Refs. \cite{zrap, zon}. 
The shape of the photon spectrum differed from the prediction,
and the total probability of the REC process per ordinary nonradiative decay was
found to be \emph{larger} by a factor of 6 than the predicted one.
To resolve this discrepancy a possibility of photon emission by the nucleus,
in so called detour transitions, was examined by Kalinowski et al. \cite{detour,kalinowski}
following ideas developed by Ford and Martin \cite{ford}.
According to Refs. \cite{detour,kalinowski, ford},
the nuclear contribution to the REC process
accompanying forbidden transitions can be substantial. In particular, for the
case of $^{41}$Ca it was claimed \cite{detour, kalinowski}
that the detour transitions fully account
for the missing intensity established by the experiment. However, for another
first forbidden unique transition --- the 1S EC decay of $^{204}$Tl ---
a different situation was encountered. The measured intensity of the REC
spectrum \cite{kurcewicz} was found to be \emph{smaller} by a factor of 4
than the value predicted by the model of Zon leaving no room for the nuclear
contribution, in contradiction with results of  Ref. \cite{kalinowski}.

Although we agree in general with Refs. \cite{ford,kalinowski} that nuclear
radiation takes place, we point out that the separation of radiation emitted
by the electron and the nucleus, respectively, is not physical because
it depends on the particular gauge used in the description of the electromagnetic field.
We argue, that although physical results do not depend on the selected gauge,
the so called \emph{length} gauge is preferred for the actual calculations.
First, it suppresses the nuclear contribution, and second, it makes possible
important simplifications in the calculations. We note that some formulae of
Zon in \cite{zon} for $\Delta J=2$ transition are divergent for the
point nucleus. It means that approximations which lead to these formulae may
not be correct. For example, an assumption that the radius of the region where
the photon emission occurs is much larger than the dimension of the
nucleus where the capture takes place, is not valid in the Coulomb
(velocity) gauge used by the authors of Refs.~\cite{zrap,zon,kalinowski}.

The failure of the theory of Zon and Rapoport to describe the experimental
data for forbidden EC transitions, and the controversy concerning the nuclear
contribution to the REC process, motivated us to address these questions
theoretically in an independent and different way.
In this work we recalculate the radiative electron capture process in
the length gauge. We restrict ourselves to the 1S electron (K-capture) only,
but the extension to other states is straightforward.
In the following section we describe our calculations in detail.
The results, in terms of a dimensionless shape factor, for the case of the first forbidden
unique transitions are presented in Section III and are compared to the experimental data
for $^{41}$Ca and $^{204}$Tl. Comparison to previous calculations is made
in Section IV followed by a summary in Section V.

\section{Theory}
\subsection{Preliminaries}
In the following we use natural units $\hbar=c=1$, $e^2 = 4\,\pi\,\alpha$ and
set the electron mass $m=1$. The Hamiltonian for the EC-decay is
\begin{equation}
H_{EC} = \frac{G_F}{\sqrt{2}}\,
          [\bar \psi_n\,\gamma_\mu\,(1-\lambda\,\gamma^5)\,\psi_p]\;
          [\bar \psi_\nu\,\gamma^\mu\,(1-\gamma^5)\,\psi_e]\,,
\end{equation}
where $\psi_n$, $\psi_p$, $\psi_\nu$ and $\psi_e$ denote neutron, proton, neutrino and electron
bispinors respectively, with Fermi constant $G_F = 1.166\,39(1)\,10^{-5}$ GeV$^{-2}$
and $\lambda = 1.26992(69)$. We use convention of Bj\"orken and Drell
\cite{bd} for Dirac matrices with
$\gamma^5 = \gamma_5 = i\,\gamma^0\,\gamma^1\,\gamma^2\,\gamma^3$, namely
\begin{equation}
\gamma^0 = \biggl(\begin{array}{cc}
I&0\\
0&-I
\end{array}\biggr),\;
\vec\gamma = \biggl(\begin{array}{cc}
0&\vec\sigma\\
-\vec\sigma&0
\end{array}\biggr),\;
\gamma^5 = \biggl(\begin{array}{cc}
0&I\\
I&0
\end{array}\biggr)\,,
\end{equation}
and $\vec\alpha = \gamma^0\,\vec\gamma$.

It is convenient to use states with a definite angular momentum.
For this reason we introduce spin spherical harmonics $\chi_\kappa^m$ \cite{edmonds}
\begin{equation}
\chi_\kappa^m(\theta,\phi) =
\left(\begin{array}{c}
\frac{-\kappa}{|\kappa|}\,\sqrt{\frac{\kappa+1/2-m}{2\,\kappa+1}}\,Y^{m-1/2}_{|\kappa+1/2|-1/2}\\
\sqrt{\frac{\kappa+1/2+m}{2\,\kappa+1}}\,Y^{m+1/2}_{|\kappa+1/2|-1/2}
\end{array}\right)\,,
\end{equation}
where
\begin{equation}
\kappa = \left\{\begin{array}{cc}
   j+1/2 &{\rm for}\, j=l-1/2\,,\\
-(j+1/2) &{\rm for}\, j=l+1/2\,.
\end{array}\right.
\end{equation}
They have the following properties
\begin{eqnarray}
\vec\sigma\cdot\hat r \chi_\kappa^m &=& -\chi_{-\kappa}^m\,,\\
\vec\sigma\cdot\vec L \chi_\kappa^m &=& -(\kappa+1)\,\chi_\kappa^m\,.
\end{eqnarray}
which are used to solve the Dirac equation. We consider at first the left handed
neutrino,
\begin{equation}
\frac{1}{2}\bigl(1-\gamma^5\bigr)\,\psi_\nu = \psi_\nu\,.
\end{equation}
Its  wave function $\psi_\nu$ with the angular momentum $j$ and the energy $q$ is
\begin{eqnarray}
\psi_\nu(\vec r) &=& \left(\begin{array}{c}\nu\\-\nu\end{array}\right)\,,\\
\nu &=& q\,[j_{j-1/2}(q\,r)\,\chi^m_{-(j+1/2)}(\theta,\phi) +
i\,j_{j+1/2}(q\,r)\,\chi^m_{j+1/2}(\theta,\phi)]\,,
\end{eqnarray}
with $j_l(x)$ being the spherical Bessel function and $\nu$ solution of
$\vec\sigma\cdot\vec p\,\nu = -q\,\nu$. The above neutrino wave function
is normalized with respect to the energy, namely
\begin{equation}
\int d^3 r\,\psi_\nu^+(\vec r)\,\psi_\nu(\vec r) = 2\,\pi\,\delta(q-q')\,.
\end{equation}

The electron wave function is the solution of the Dirac equation in the Coulomb
field $-Z\,\alpha/r$ \cite{dhbk}. It is of the form
\begin{equation}
\psi_e(\vec r) = \left(\begin{array}{c}
G(r)\,\chi^m_\kappa(\theta,\phi)\\
i\,F(r)\, \chi^m_{-\kappa}(\theta,\phi)
\end{array}\right)\,, \label{11}
\end{equation}
with the energy $E=E_{n,\kappa}$
\begin{equation}
E_{n,\kappa} = \left(1+\frac{Z^2\,\alpha^2}{(n+1+\gamma)^2}\right)^{-1/2}\,,
\end{equation}
where $\gamma = -1+\sqrt{\kappa^2-Z^2\,\alpha^2}$. The explicit form of radial
wave functions $G$ and $F$ is \cite{dhbk}
\begin{eqnarray}
G_{n,\kappa}(r) &=& C_{n,\kappa}\,(2\,\lambda\,r)^\gamma\,e^{-\lambda\,r}\,
\sqrt{1+E_{n,\kappa}}\;[g^{(2)}_{n,\kappa}(r) + g^{(1)}_{n,\kappa}(r)]\,, \\
F_{n,\kappa}(r) &=& C_{n,\kappa}\,(2\,\lambda\,r)^\gamma\,e^{-\lambda\,r}\,
\sqrt{1-E_{n,\kappa}}\;[g^{(2)}_{n,\kappa}(r) - g^{(1)}_{n,\kappa}(r)]\,,\\
g^{(1)}_{n,\kappa}(r) &=&
\sqrt{Z\,\alpha\,\lambda^{-1}-\kappa}\;L_n^{(2+2\,\gamma)}(2\,\lambda\,r)\,,\\
g^{(2)}_{n,\kappa}(r) &=& -(n+2+2\,\gamma)\,\sqrt{Z\,\alpha\,\lambda^{-1}-\kappa}\;
L_{n-1}^{(2+2\,\gamma)}(2\,\lambda\,r)\,,\\
C_{n,\kappa} &=& \sqrt{\frac{2\,\lambda^4\,n!}{Z\,\alpha\,\Gamma(n+3+2\,\gamma)}}\,,
\end{eqnarray}
where $L_n^{(\alpha)}$ are Laguerre polynomials
\begin{equation}
L_n^{(\alpha)}(z) =
\sum_{k=0}^{n} \frac{\Gamma(\alpha+n+1)}{k!\,(n-k)!\,\Gamma(\alpha+k+1)}\,(-z)^k\,,
\end{equation}
with $L_{-1}\equiv0$ and $\lambda = \sqrt{1-E^2}$. In particular, the energy and
the wave function of the ground state $n=l=0,\kappa=-1$, and $j=1/2$ are given by
\begin{eqnarray}
E_{0,-1} &=& \sqrt{1-Z^2\,\alpha^2} \equiv {\cal E}\,,\\
G_{0,-1} &=& \sqrt{\frac{4\,Z^3\,\alpha^3\,(2+\gamma)}{\Gamma(3+2\,\gamma)}}\,
(2\,Z\,\alpha\,r)^\gamma\,e^{-Z\,\alpha\,r}\equiv {\cal G}\,, \label{20}\\
F_{0,-1} &=& -\sqrt{\frac{4\,Z^3\,\alpha^3\,(-\gamma)}{\Gamma(3+2\,\gamma)}}\,
(2\,Z\,\alpha\,r)^\gamma\,e^{-Z\,\alpha\,r}\equiv {\cal F}\,. \label{21}
\end{eqnarray}
For the calculation of radiative transition rates one needs the
Dirac-Coulomb Green's function $G^D$. Its explicit form is \cite{dhbk}
\begin{eqnarray}
G^D(\vec r,\vec r\,',E) &\equiv& \langle \vec r\,|\frac{1}{H-E}|\vec r\,'\rangle
= \sum_{\kappa m}G^D_{\kappa,m}(\vec r,\vec r\,',E)\,, \label{22}\\
G^{D}_{\kappa,m}(\vec r,\vec r\,',E) &=& 
\Theta(r'-r)\,\psi^<_{\kappa m}(\vec r\,)\otimes \psi^>_{\kappa m}(\vec r\,')^\dagger +
\Theta(r-r')\,\psi^>_{\kappa m}(\vec r\,)\otimes \psi^<_{\kappa m}(\vec r\,')^\dagger\,, \label{23}
\end{eqnarray}
where $\Theta(x)$ is a Heaviside step function,
and both $\psi^>$ and $\psi^<$ are of the form (\ref{11}) with
\begin{eqnarray}
G_\kappa^<(r) &=& (2\,\lambda\,r)^\gamma\,e^{-\lambda\,r}\,\sqrt{1+E}\,(f_2+f_1)\,,\\
F_\kappa^<(r) &=& (2\,\lambda\,r)^\gamma\,e^{-\lambda\,r}\,\sqrt{1-E}\,(f_2-f_1)\,,\\
G_\kappa^>(r) &=&
(2\,\lambda\,r)^\gamma\,e^{-\lambda\,r}\,\sqrt{1+E}\,(f_4+f_3)\,
\frac{2\,\lambda\,\Gamma(a)}{\Gamma(c)}\,,\\
F_\kappa^>(r) &=& (2\,\lambda\,r)^\gamma\,e^{-\lambda\,r}\,\sqrt{1-E}\,(f_4-f_3)\,
\frac{2\,\lambda\,\Gamma(a)}{\Gamma(c)}\,,
\end{eqnarray}
where
\begin{eqnarray}
f_1 &=& (Z\,\alpha\,\lambda^{-1}-\kappa)\,_1F_1(a,c,2\,\lambda\,r)\,,\\
f_2 &=& a\,_1F_1(a+1,c,2\,\lambda\,r)\,,\\
f_3 &=& U(a,c,2\,\lambda\,r)\,,\\
f_4 &=& (Z\,\alpha\,\lambda^{-1}+\kappa)\,U(a+1,c,2\,\lambda\,r)\,, \label{31}
\end{eqnarray}
and $a = 1+\gamma-E\,Z\,\alpha\,\lambda^{-1}$, $c=3+2\,\gamma$,
while $_1F_1$ and $U$ are confluent hypergeometric functions
regular at the origin and at the infinity, respectively.

To describe photon wave function we introduce vector spherical harmonics
\cite{edmonds},
\begin{equation}
\vec Y_{JL}^M(\theta,\phi) = \sum_{m\,q}\,Y_L^m(\theta,\phi)\,\vec e_q\,
\langle L\,m;1\,q|L\,1;J\,M\rangle\,,
\end{equation}
where $\vec e_1 = -1/\sqrt{2}\,(\vec e_x+i\,\vec e_y)$, $\vec e_0 = \vec e_z$,
and $\vec e_{-1} = 1/\sqrt{2}\,(\vec e_x-i\,\vec e_y)$.
The solutions of the Maxwell equations with definite angular momentum and parity
are represented by both, the magnetic photon
\begin{eqnarray}
\vec A^{\,(\!M)}_{JM}(\vec r) &=& \sqrt{2\,k}\,j_J(k\,r)\,
\vec Y_{JJ}^M(\hat r) \label{33}\\
A^{0\,(\!M)}_{JM}(\vec r) &=& 0
\end{eqnarray}
and the electric photon in the length gauge \cite{akhiezer}
\begin{eqnarray}
\vec A^{\,(\!E)}_{JM}(\vec r) &=& \sqrt{2\,k}\,\sqrt{\frac{2\,J+1}{J}}\,j_{J+1}(k\,r)\,
\vec Y_{JJ+1}^M(\hat r)\,, \label{35}\\
A^{0\,(\!E)}_{JM}(\vec r) &=& -i\,\sqrt{2\,k}\,\sqrt{\frac{J+1}{J}}\,j_J(k\,r)\,
Y_J^M(\hat r)\,,\label{36}
\end{eqnarray}
while the electric photon in the Coulomb (velocity) gauge is
\begin{eqnarray}
\vec A^{\,(\!E)}_{JM}(\vec r) &=& \sqrt{2\,k}\,\left[
\sqrt{\frac{J}{2\,J+1}}\,j_{J+1}(k\,r)\,\vec Y_{JJ+1}^M(\hat r)-
\sqrt{\frac{J+1}{2\,J+1}}\,j_{J-1}(k\,r)\,\vec Y_{JJ-1}^M(\hat r)
\right], \label{35p}\\
A^{0\,(\!E)}_{JM}(\vec r) &=& 0\,.\label{36p}
\end{eqnarray}
These solutions are normalized with respect to energy, so that
\begin{equation}
2\,k\,\int d^3r\,[\vec A_{JM}(\vec r)^*\;\vec A'_{JM}(\vec r)
-A^{0}_{JM}(\vec r)^*\;A'^0_{JM}(\vec r)]
= 2\,\pi\,\delta(k-k')\,,
\end{equation}
The use of the length gauge is essential for performing
several simplifications in the calculations of transition
rates, as it is explained in the next sections.

\subsection{Electron capture rate}
The electron capture rate $W$ is equal to the square of the the matrix element
$M$, summed over final states and averaged over initial states
\begin{equation}
W = \frac{1}{2\,J_e+1}\,\frac{1}{2\,J_i+1}\,\sum_{M_e\,M_i\,M_\nu\,M_f}\,|M|^2\,,
\end{equation}
where
\begin{equation}
M = \langle f|H_{EC}|i\rangle = \int d^3 r
\frac{G_F}{\sqrt{2}}\,
          [\bar \psi_f\,\gamma_\mu\,(1-\lambda\,\gamma^5)\,\psi_i]\;
          [\bar \psi_\nu\,\gamma^\mu\,(1-\gamma^5)\,\psi_e]\,. \label{39}
\end{equation}
Although we use single nucleon matrix elements, results can easily be
transformed for the nuclear matrix elements, by assuming that
$\psi_n$ and $\psi_p$ are field operators, and instead of
$\bar \psi_f\,\gamma_\mu\,(1-\lambda\,\gamma^5)\,\psi_i$
one considers $\langle f|\bar \psi_n\,\gamma_\mu\,(1-\lambda\,\gamma^5)\,\psi_p|i\rangle$.
The tensor decomposition of the matrix element in Eq. (\ref{39}) leads to
\begin{equation}
M = \frac{G_F}{\sqrt{2}}\,\int r^2 dr\,\sum_{JLSM} (-1)^{J+M}\,
\bigl(f|T^M_{JLS}\,(1-\lambda\,\gamma^5)|i\bigr)\,
\bigl(\psi_\nu|T^{-M}_{JLS}\,(1-\gamma^5)|\psi_e\bigr)\,,
\end{equation}
where
\begin{eqnarray}
T^M_{JL0} &=& i^L\,\delta_{JL}\,Y_L^M\,,\\
T^M_{JL1} &=& (-1)^{J+L+1}\,i^L\,\vec Y^M_{JL}\cdot\vec\alpha\,,
\end{eqnarray}
and $(.|.)$ denotes the integral over angular coordinates.
Each state $f,i,\psi_\nu$ and $\phi_e$ has definite angular momentum
$J,M$ numbers, so one can use reduced matrix element
\begin{equation}
\langle j,m|T^q_k|j',m'\rangle = (-1)^{j-m}\,
\left(\begin{array}{ccc} j&k&j'\\
                         -m&q&m' \end{array}\right)\,\langle j||T_k||j'\rangle\,,
\end{equation}
and orthogonality properties of $3j$ symbol \cite{edmonds}
to obtain simple formula for the electron capture rate
\begin{eqnarray}
W &=& \frac{G_F^2}{2}\,\frac{1}{2\,J_e+1}\,\frac{1}{2\,J_i+1}\,
\sum_J\,\frac{1}{2\,J+1}\nonumber \\&&\times
\biggl|\int r^2
dr\,\sum_{LS}\,\bigl(J_f||T_{JLS}\,(1-\lambda\,\gamma^5)||J_i\bigr)\,
\bigl(J_\nu||T_{JLS}\,(1-\gamma^5)||J_e\bigr)\biggr|^2\,.
\end{eqnarray}
The reduced matrix elements of spherical harmonics
are given by
\begin{eqnarray}
\langle\kappa_f||Y_l||\kappa_i\rangle &=&
\sqrt{\frac{2\,l+1}{4\,\pi}}\,C_l(\kappa_f,\kappa_i)\,, \\
C_l(\kappa_f,\kappa_i) &=&
(-1)^{j_f+1/2}\,\sqrt{(2\,j_f+1)(2\,j_i+1)}\,\left(
\begin{array}{ccc}
j_i&l&j_f\\ 1/2&0&-1/2\end{array}\right)\,\Pi(l_f,l_i,l)\,,\\
\Pi(l_f,l_i,l) &=& \frac{1}{2}\,\bigl[1+(-1)^{l_f+l_i+l}\bigr]\,,
\end{eqnarray}
and spin spherical harmonics by
\begin{eqnarray}
\langle\kappa_f||\vec Y_{JL}\cdot\vec\sigma||\kappa_i\rangle &=&
\sqrt{\frac{2\,J+1}{4\,\pi}}\,S_{JL}(\kappa_f,\kappa_i)\,, \\
S_{J,J+1}(\kappa_f,\kappa_i) &=&
\sqrt{\frac{J+1}{2\,J+1}}\,\biggl(1+\frac{\kappa_f+\kappa_i}{J+1}\biggr)\,
C_J(-\kappa_f,\kappa_i)\,, \\
S_{J,J}(\kappa_f,\kappa_i) &=& \frac{\kappa_i-\kappa_f}{\sqrt{J\,(J+1)}}\,C_J(\kappa_f,\kappa_i)\,, \\
S_{J,J-1}(\kappa_f,\kappa_i) &=&
\sqrt{\frac{J}{2\,J+1}}\,\biggl(-1+\frac{\kappa_f+\kappa_i}{J}\biggr)\,
C_J(-\kappa_f,\kappa_i)\,.
\end{eqnarray}
With the use of the above formulae for reduced matrix elements,
the capture rate $W$ of the 1S electron by the nucleus is
\begin{eqnarray}
W &=& \sum_J W_J\,,  \\ \nonumber \\
W_0 &=& \frac{2\,G_F^2\,Q^2}{2\,J_i+1}\,\frac{1}{4\,\pi}\,
\bigl|\langle J_f||\bigl[T_{000}\,{\cal G}(R)
-T_{011}\,{\cal F}(R)\bigr]\,(1-\lambda\,\gamma_5)||J_i\rangle\bigr|^2\,,\\
\nonumber \\
W_1 &=&\frac{2\,G_F^2\,Q^2}{2\,J_i+1}\,\frac{1}{4\,\pi}\,
\bigl|\langle J_f||\bigl[T_{101}\,{\cal G}(R)
+\bigl(T_{110}/\sqrt{3}+T_{111}\,\sqrt{2/3}\bigr)
\,{\cal F}(R)\bigr]\,(1-\lambda\,\gamma_5)||J_i\rangle\bigr|^2\,,
\nonumber \\ \\
W_2 &=&\frac{2}{9}\,\frac{G_F^2\,Q^4}{2\,J_i+1}\,\frac{1}{4\,\pi}\,
\bigl|\langle J_f||r\,\bigl[T_{211}\,{\cal G}(R)
+\bigl(T_{220}\sqrt{2/5}+T_{221}\sqrt{3/5}\bigr)
{\cal F}(R)\bigr](1-\lambda\,\gamma_5)||J_i\rangle\bigr|^2 \,,
\nonumber \\
\end{eqnarray}
where $Q$ is the energy released in the decay and $R$ is the nuclear radius.

\subsection{Radiative electron capture rate}
The probability amplitude for the electron capture with the simultaneous photon emission is
\begin{equation}
M_R = \int d^3 r
\frac{G_F}{\sqrt{2}}\,
          [\bar \psi_f\,\gamma_\mu\,(1-\lambda\,\gamma^5)\,\psi_i]\;
          [\bar \psi_\nu\,\gamma^\mu\,(1-\gamma^5)\,\psi'_e]\,,
\end{equation}
where
\begin{equation}
\psi'_e(\vec r) = \langle \vec r|\frac{1}{{\cal E}-H-k}\,e(A^0-\vec\alpha\cdot\vec A)|\psi_e\rangle
=-\sum_n \psi'_{n}(r)\,\frac{\langle\psi'_{n}|e(A^0-\vec\alpha\cdot\vec
 A)|\psi_e\rangle}{E_n+k-{\cal E}}\,.
\end{equation}
We use the latter form to perform tensor decomposition of $M_R$ and obtain
\begin{eqnarray}
M_R &=& \frac{G_F\,e}{\sqrt{2}}\,\int r^2 dr\,\sum_{JLSMM'_e} (-1)^{J+M+1}\,
\bigl(J_f,M_f|T^M_{JLS}\,(1-\lambda\,\gamma^5)|J_i,M_i\bigr)\,\sum_{n}\\ &&
\bigl(J_\nu,M_\nu|T^{-M}_{JLS}\,(1-\gamma^5)|n,J'_e,M'_e\bigr)\,
\frac{1}{E_n+k-{\cal E}}\,\langle n,J'_e,M'_e|
A^0_{J_A M_A}-\vec\alpha\cdot\vec A_{J_A M_A}|J_e,M_e\rangle \nonumber
\end{eqnarray}
The rate for the radiative electron capture $W_R$ is
\begin{equation}
W_R = \int_0^Q dk\;W_R(k)\,,
\end{equation}
where
\begin{equation}
W_R(k) = \frac{1}{2\,J_e+1}\,\frac{1}{2\,J_i+1}\,
\sum_{M_e M_i M\nu M_f M_A}\,\frac{1}{2\,\pi}\,|M_R|^2\,,
\end{equation}
with the sum over final state and the average over initial states.
The $W_R(k)$ can be expressed in terms of the reduced matrix elements and
the summation over magnetic states can be carried out with the aid
of the orthogonality of the $3j$ symbols. The results is
\begin{eqnarray}
W_R(k) &=&
G_F^2\,\alpha\,\frac{1}{2\,J_e+1}\,\frac{1}{2\,J_i+1}\,\sum_J\,\frac{1}{2\,J+1}\,
\sum_{J'_e}\,\frac{1}{2\,J'_e+1}\,\Bigl|\sum_{n}\frac{1}{E_n+k-{\cal E}}\nonumber
\\ &&
\times \int r^2 dr \sum_{LS}\,\bigl(J_f||T_{JLS}(1-\lambda\,\gamma^5)||J_i\bigr)\,
\bigl(J_\nu||T_{JLS}(1-\gamma^5)||J'_e,n\bigr) \nonumber \\ && \times
\langle n,J'_e||A^0_{J_A}-\vec\alpha\cdot\vec A_{J_A}||J_e\rangle\Bigr|^2\,. \label{64}
\end{eqnarray}
We can use now the explicit form of the Dirac-Coulomb Green's function in
Eq. (\ref{22}-\ref{31}) to replace the sum over intermediate electron states
{\em n} in Eq. (\ref{64}) by $G^D$. Since the electron capture takes place within the nucleus,
and the photon radiation in a region of the electron wave function which is
several orders of magnitude larger, we apply an identity
$\Theta(r'-r) = 1-\Theta(r-r')$
and neglect $\Theta(r-r')$ in Eq. (\ref{23}) completely, so the Green's
function becomes
\begin{equation}
G^{D}_{\kappa,m}(\vec r,\vec r\,',E) \approx 
\psi^<_{\kappa m}(\vec r\,)\otimes \psi^>_{\kappa m}(\vec r\,')^\dagger\,. \label{65}
\end{equation}
In other words, this approximation is allowed, because the integral with $\Theta(r-r')$
gives contribution, which is higher order in the small parameter $\xi = Q\,R$.
After this assumption one obtains
\begin{eqnarray}
W_R(k) &=&
\frac{G_F^2}{2}\,\frac{1}{2\,J_i+1}\,\sum_J\,\frac{1}{2\,J+1}\,
\sum_{J'_e}\,\frac{1}{2\,J'_e+1}
\nonumber \\ &&
\times \Bigl|\int r^2 dr \sum_{LS}\,\bigl(J_f||T_{JLS}(1-\lambda\,\gamma^5)||J_i\bigr)\,
\bigl(J_\nu||T_{JLS}(1-\gamma^5)||J'_e\,\psi^<\bigr)\Bigr|^2\nonumber \\ &&
\times\biggl\{\frac{2\,\alpha}{2\,J_e+1}\,
\bigl|\langle J'_e,\psi^>||A^0_{J_A}-\vec\alpha\cdot\vec
A_{J_A}||J_e\rangle\bigr|^2\biggr\}\,.
\label{63}
\end{eqnarray}
The approximation in Eq. (\ref{65}) would not be valid in the Coulomb gauge,
as the integral with $\Theta(r-r')$ is of the same order 
in the parameter $\xi$, what we discuss in more details in Sec. IV.

We consider transitions with $\Delta J=0,1,2$
\begin{equation}
W_R(k) = W_{R0}(k) +  W_{R1}(k) +  W_{R2}(k)\,,
\end{equation}
since no electron radiative capture has been
observed for higher multipolarities.
Due to the assumption in Eq. (\ref{65}), transition rate
for each value of $J$ can be decomposed into a product
of a term corresponding to the photon emission and
a term corresponding to the nuclear transition, namely
\begin{eqnarray}
W_{R0}(k) &=& \;\;W_{M1}(k, S_{1/2}\rightarrow S_{1/2}^>)\,
             W_0(Q-k,S_{1/2}^<\rightarrow \nu_{1/2}) \nonumber \\ &&
            +W_{E1}(k, S_{1/2}\rightarrow P_{1/2}^>)\,
             W_0(Q-k,P_{1/2}^<\rightarrow \nu_{1/2})\,,\\ \nonumber \\
W_{R1}(k) &=& \;\;W_{M1}(k, S_{1/2}\rightarrow S_{1/2}^>)\,
             W_1(Q-k,S_{1/2}^<\rightarrow \nu_{1/2}) \nonumber \\ &&
            +W_{E1}(k, S_{1/2}\rightarrow P_{1/2}^>)\,
             W_1(Q-k,P_{1/2}^<\rightarrow \nu_{1/2})\,,\\\nonumber \\
W_{R2}(k) &=& \;\;W_{M1}(k, S_{1/2}\rightarrow S_{1/2}^>)\,
             W_2(Q-k,S_{1/2}^<\rightarrow \nu_{3/2}) \nonumber \\ &&
            +W_{E1}(k, S_{1/2}\rightarrow P_{1/2}^>)\,
             W_2(Q-k,P_{1/2}^<\rightarrow \nu_{3/2}) \\ &&
            +\bigl[W_{E1}(k, S_{1/2}\rightarrow P_{3/2}^>) +W_{M2}(k, S_{1/2}\rightarrow P_{3/2}^>)\bigr]\,
             W_2(Q-k,P_{3/2}^<\rightarrow \nu_{1/2}) \nonumber \\ &&
            +\bigl[W_{M1}(k, S_{1/2}\rightarrow D_{3/2}^>)+ W_{E2}(k, S_{1/2}\rightarrow D_{3/2}^>)\bigr]\,
             W_2(Q-k,D_{3/2}^<\rightarrow \nu_{1/2})\,. \nonumber
\end{eqnarray}
The explicit formulae for the photon emission rate
\begin{equation}
W_{EM} = \frac{2\,\alpha}{2\,J_e+1}\,
\bigl|\langle J'_e,\psi^>||A^0_{J_A}-\vec\alpha\cdot\vec
A_{J_A}||J_e\rangle\bigr|^2
\equiv \frac{\alpha}{\pi}\,k\,{\cal R}_{EM}
\end{equation}
are obtained using
photon wave function from Eq. (\ref{33}-\ref{36}), electron wave function from Eqs. (\ref{20},\ref{21}),
and the Dirac-Coulomb Green's function from Eq. (\ref{65})
\begin{eqnarray}
{\cal R}_{M1}(S_{1/2}\rightarrow S_{1/2}^>) &=&
2\,\Bigl|\int r^2\,dr\,j_1(k\,r)\,\bigl[G^>_{-1}({\cal E}-k,r)\,{\cal F}(r)
+F^>_{-1}({\cal E}-k,r)\,{\cal G}(r)\bigr]\Bigr|^2\,,\nonumber \\ \\
{\cal R}_{M1}(S_{1/2}\rightarrow D_{3/2}^>) &=&
\Bigl|\int r^2\,dr\,j_1(k\,r)\,\bigl[G^>_{2}({\cal E}-k,r)\,{\cal F}(r)
+F^>_{2}({\cal E}-k,r)\,{\cal G}(r)\bigr]\Bigr|^2 \,,\nonumber \\ \\
{\cal R}_{M2}(S_{1/2}\rightarrow P_{3/2}^>) &=&
3\,\Bigl|\int r^2\,dr\,j_2(k\,r)\,\bigl[G^>_{-2}({\cal E}-k,r)\,{\cal F}(r)
+F^>_{-2}({\cal E}-k,r)\,{\cal G}(r)\bigr]\Bigr|^2 \,,\nonumber \\ \\
{\cal R}_{E1}(S_{1/2}\rightarrow P_{1/2}^>) &=&
2\,\Bigl|\int r^2\,dr\,\Bigl\{j_1(k\,r)\,\bigl[G^>_{1}({\cal E}-k,r)\,{\cal G}(r)
+F^>_{1}({\cal E}-k,r)\,{\cal F}(r)\bigr] \nonumber \\ && + 2\,j_2(k\,r)\,G^>_1({\cal E}-k,r)\,{\cal F}(r)\Bigr\}\Bigr|^2\,, \\
{\cal R}_{E1}(S_{1/2}\rightarrow P_{3/2}^>) &=&
\Bigl|\int r^2\,dr\,\Bigl\{2\,j_1(k\,r)\,\bigl[G^>_{-2}({\cal E}-k,r)\,{\cal G}(r)+F^>_{-2}({\cal E}-k,r)\,{\cal F}(r)\bigr]
\nonumber\\ &&
+j_2(k\,r)\,\bigl[G^>_{-2}({\cal E}-k,r)\,{\cal F}(r)-3\,F^>_{-2}({\cal E}-k,r)\,{\cal G}(r)\bigr]\Bigr\}\Bigr|^2\,, \label{75}\\
{\cal R}_{E2}(S_{1/2}\rightarrow D_{3/2}^>) &=&
3\,\Bigl|\int r^2\,dr\,\Bigl\{j_2(k\,r)\,\bigl[G^>_{2}({\cal E}-k,r)\,{\cal G}(r)
+F^>_{2}({\cal E}-k,r)\,{\cal F}(r)\bigr]\nonumber \\ && + 2\,j_3(k\,r)\,G^>_2({\cal E}-k,r)\,{\cal F}(r)\Bigr\}\Bigr|^2\,,
\end{eqnarray}
where $E = E_{0,-1}$, ${\cal F}(r) = F_{0,-1}(r)$, and ${\cal G}(r) = G_{0,-1}(r)$. The
explicit formulae for the electron capture rate are
\begin{eqnarray}
W_0(S_{1/2}^<\rightarrow \nu_{1/2}) &=& \frac{2\,G_F^2\,(Q-k)^2}{2\,J_i+1}\,\frac{1}{4\,\pi}\,
\bigl|\langle J_f|| \bigl[T_{000}\,G^<_{-1}({\cal E}-k,R)
\nonumber \\ &&
-T_{011}\,F^<_{-1}({\cal E}-k,R)\bigr]\,(1-\lambda\,\gamma_5)||J_i\rangle\bigr|^2\,,\\
\nonumber \\
W_0(P_{1/2}^<\rightarrow \nu_{1/2}) &=& \frac{2\,G_F^2\,(Q-k)^2}{2\,J_i+1}\,\frac{1}{4\,\pi}\,
\bigl|\langle J_f|| \bigl[T_{000}\,F^<_{1}({\cal E}-k,R)
\nonumber \\ &&
+T_{011}\,G^<_{1}({\cal E}-k,R)\bigr]\,(1-\lambda\,\gamma^5)||J_i\rangle\bigr|^2\,,\\
\nonumber \\
W_1(S_{1/2}^<\rightarrow \nu_{1/2}) &=&\frac{2\,G_F^2\,(Q-k)^2}{2\,J_i+1}\,\frac{1}{4\,\pi}\,
\bigl|\langle J_f|| \bigl[T_{101}\,G^<_{-1}({\cal E}-k,R)
\nonumber \\ &&
+\bigl(T_{110}/\sqrt{3}+T_{111}\,\sqrt{2/3}\bigr)\,
F^<_{-1}({\cal E}-k,R)\bigr]\,(1-\lambda\,\gamma^5)||J_i\rangle\bigr|^2\,,
 \\ \nonumber \\
W_1(P_{1/2}^<\rightarrow \nu_{1/2}) &=&\frac{2\,G_F^2\,(Q-k)^2}{2\,J_i+1}\,\frac{1}{4\,\pi}\,
\bigl|\langle J_f|| \bigl[T_{101}\,F^<_{1}({\cal E}-k,R)
\nonumber \\ &&
-\bigl(T_{110}/\sqrt{3} + T_{111}\,\sqrt{2/3}\bigr)\,G^<_{1}({\cal E}-k,R)
 \bigr]\,(1-\lambda\,\gamma^5)||J_i\rangle\bigr|^2\,,
 \\ \nonumber \\
W_2(S_{1/2}^<\rightarrow \nu_{3/2}) &=&\frac{2}{9}\,\frac{G_F^2\,(Q-k)^4}{2\,J_i+1}\,\frac{1}{4\,\pi}\,
\bigl|\langle J_f|| r\,\bigl[T_{211}\,G^<_{-1}({\cal E}-k,R)
\nonumber \\ &&
+\bigl(T_{220}\,\sqrt{2/5} + T_{221}\,\sqrt{3/5}\bigr)\,F^<_{-1}({\cal E}-k,R)
\bigr]\,(1-\lambda\,\gamma^5)||J_i\rangle\bigr|^2\,,
 \nonumber \\ \\
W_2(P_{1/2}^<\rightarrow \nu_{3/2}) &=&\frac{2}{9}\,\frac{G_F^2\,(Q-k)^4}{2\,J_i+1}\,\frac{1}{4\,\pi}\,
\bigl|\langle J_f|| r\,\bigl[T_{211}\,F^<_{1}({\cal E}-k,R)
\nonumber \\ &&
-\bigl(T_{220}\,\sqrt{2/5} + T_{221}\,\sqrt{3/5}\bigr)\,G^<_{1}({\cal E}-k,R)
\bigr]\,(1-\lambda\,\gamma^5)||J_i\rangle\bigr|^2\,,
 \nonumber \\ \\
W_2(P_{3/2}^<\rightarrow \nu_{1/2}) &=& \frac{G_F^2\,(Q-k)^2}{(2\,J_i+1)\,R^2}\,\frac{1}{4\,\pi}\,
\bigl|\langle J_f|| r\,\bigl[T_{211}\,G^<_{-2}({\cal E}-k,R)
\nonumber \\ &&
+\bigl(T_{220}\,\sqrt{2/5} + T_{221}\,\sqrt{3/5}\bigr)\,F^<_{-2}({\cal E}-k,R)
\bigr]\,(1-\lambda\,\gamma^5)||J_i\rangle\bigr|^2\,,
 \nonumber \\ \\
W_2(D_{3/2}^<\rightarrow \nu_{1/2}) &=& \frac{G_F^2\,(Q-k)^2}{(2\,J_i+1)\,R^2}\,\frac{1}{4\,\pi}\,
\bigl|\langle J_f|| r\,\bigl[T_{211}\,F^<_{2}({\cal E}-k,R)
\nonumber \\ &&
-\bigl(T_{220}\,\sqrt{2/5} + T_{221}\,\sqrt{3/5}\bigr)\,G^<_{2}({\cal E}-k,R)
\bigr]\,(1-\lambda\,\gamma^5)||J_i\rangle\bigr|^2\,. \nonumber \\
\end{eqnarray}

\section{Results}
Following the generally adopted convention \cite{glauber, martin, kalinowski},
we calculate the ratio of radiative to nonradiative capture rate. Then, the
resulting photon spectrum accompanying the 1S capture is given by
\begin{equation}
\frac{W_R(k)}{W} \equiv \frac{\alpha}{\pi\,m^2}\,\biggl(1-\frac{k}{Q}\biggr)^2\,k\,{\cal R}\,,
\end{equation}
where details of the shape are expressed in terms of the dimensionless
shape factor ${\cal R}$.

First, we consider the transitions where $\Delta J=0,1$. When the parity of nuclear states
does not change, $\pi_i\,\pi_f=1$, this transition is allowed.
Then, two tensor operators $T_{000}$ and $T_{101}(-\lambda\,\gamma^5)$
dominate and other operators can be neglected.
In the probability ratio the nuclear matrix elements cancel out
and the result greatly simplifies:
\begin{equation}
{\cal R} = {\cal R}_a = \frac{G^<_{-1}({\cal E}-k,R)^2}{{\cal G}(R)^2}\,{\cal R}_{M1}(S_{1/2}\rightarrow S_{1/2}^>)
+ \frac{F^<_{1}({\cal E}-k,R)^2}{{\cal G}(R)^2}\,{\cal R}_{E1}(S_{1/2}\rightarrow P_{1/2}^>)\,. \label{84}
\end{equation}
It is worth to note, that in the nonrelativistic limit ($Z\,\alpha\rightarrow 0$),
the shape factor ${\cal R}_a$ is equal to 1, what we verified by explicit calculations.
If the parity does change, $\pi_i\,\pi_f=-1$, the transition is first
forbidden non unique. Here, all nuclear operators contribute to the transition
matrix elements and the exact expression for ${\cal R}$ in general is more complicated.
However, one notices that the ratio
\begin{eqnarray}
\frac{F^<_\kappa(E,R)}{G^<_\kappa(E,R)} =
\sqrt{\frac{1-E}{1+E}}\,\frac{\sqrt{\kappa^2-Z^2\,\alpha^2}+
\kappa-Z\,\alpha\,\sqrt{\frac{1+E}{1-E}}}
{\sqrt{\kappa^2-Z^2\,\alpha^2}-\kappa+Z\,\alpha\,\sqrt{\frac{1-E}{1+E}}}
\approx \biggl(\frac{2\,\kappa}{Z\,\alpha}\biggr)^{\kappa/|\kappa|}
\approx -\frac{G^<_{-\kappa}(E,R)}{F^<_{-\kappa}(E,R)}\,,\label{85}
\end{eqnarray}
approximately does not depend on energy $E$ as long as  $Z\,\alpha \ll 1$.
Therefore, nuclear matrix elements cancel out
and the formula in Eq. (\ref{84}) holds for non unique transitions, 
but only when this approximation is valid.
It is important to stress that numerically Eq. (\ref{84}) yields the same results as those
obtained in Refs. \cite{glauber, martin, zrap}.
Although our results have apparently different and simpler  form, they are equivalent
to those obtained previously by Glauber, Martin, Zon and Rapoport
for the allowed and first forbidden non unique decays.

The situation is different, however, in the case of the first forbidden unique transition,
where $\Delta J=2$ and $\pi_i\,\pi_f=-1$.
The operator $T_{211}(-\lambda\,\gamma^5)$ dominates and other nuclear
operators are neglected. The resulting shape factor is
\begin{eqnarray}
{\cal R}_{1u} &=& \biggl(1-\frac{k}{Q}\biggr)^2\,{\cal R}^{(1)} +
\frac{k^2}{Q^2}\,{\cal R}^{(2)}\,,\label{88}\\
{\cal R}^{(1)} &=&{\cal R}_a\,,\\
{\cal R}^{(2)} &=&\frac{9}{2\,k^2\,R^2}\,
              \frac{G^<_{-2}({\cal E}-k,R)^2}{{\cal G}(R)^2}\,
\bigl[{\cal R}_{E1}(S_{1/2}\rightarrow P_{3/2}^>)+ {\cal R}_{M2}(S_{1/2}\rightarrow P_{3/2}^>)\bigr]
\nonumber \\ &&
             +\frac{9}{2\,k^2\,R^2}\,\frac{F^<_{2}({\cal E}-k,R)^2}{{\cal G}(R)^2}\,
\bigl[{\cal R}_{M1}(S_{1/2}\rightarrow D_{3/2}^>) + {\cal R}_{E2}(S_{1/2}\rightarrow D_{3/2}^>)\bigr]\,.
\label{90}
\end{eqnarray}
While the nonrelativistic limit of ${\cal R}^{(1)}$ is $1$,
for the ${\cal R}^{(2)}$ function we obtain:
\begin{equation}
\lim_{Z\rightarrow 0} {\cal R}^{(2)} =  1+\frac{1}{k}+\frac{2}{k^2}\,,
\end{equation}
what we have verified numerically.
Although the result of Zon \cite{zon} has similar form to Eq. (\ref{88}),
the Coulomb-free limit of the function corresponding to
${\cal R}^{(2)}$ is equal to $1$, which differs significantly from our result.

\begin{figure}[h]%
\begin{center}\includegraphics[width=0.9\linewidth]{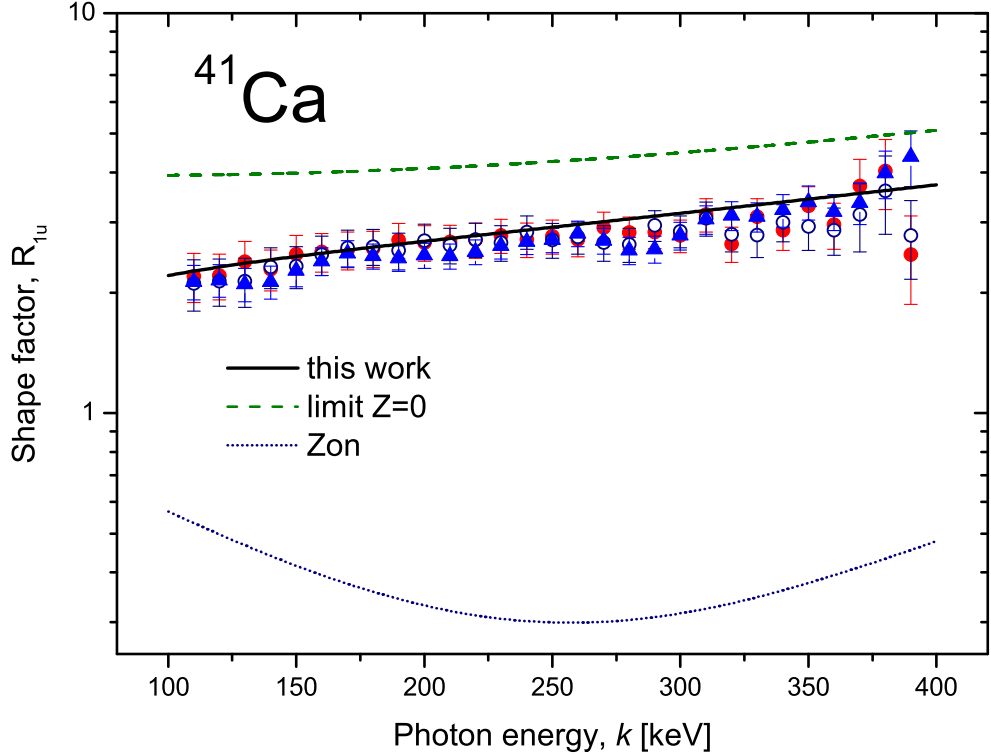}\end{center}%
\caption{\label{fig1}
The shape factor for the 1S component of the first forbidden unique REC in $^{41}$Ca.
The points with error bars represent the experimental values form Ref.\cite{41Ca}.
The solid line shows the result of this work as given by Eqs. 86-88, while the
nonrelativistic limit is shown by the dashed line. The dotted line represents
the prediction of the Zon and Rapoport \cite{zrap,zon}. The value of $Q$ is
$416.4$ keV. }
\end{figure}

Experimentally, the REC process in the first forbidden unique transition was studied
so far for two nuclei: $^{41}$Ca \cite{41Ca} and $^{204}$Tl \cite{kurcewicz}.
The data obtained for $^{41}$Ca ($Q_{EC} = 421.3$ keV, $T_{1/2} = 10^5$ years)
are shown in Fig.~1 together with theoretical predictions. While the model of
Zon and Rapoport \cite{zrap,zon} underestimates the intensity of the REC spectrum
by a factor of about 6, the results of our calculations are in excellent agreement
with the experiment.
The contribution of relativistic and Coulomb effects is evident. The Coulomb-free limit
of our result would overestimate the intensity approximately by a factor of 2.
The discrepancy between the Zon model and the $^{41}$Ca data was used as an argument
in favor of nuclear contribution to the REC process \cite{detour,kalinowski}.
In fact, if both, the nuclear and the electron radiation were calculated correctly,
the result should agree with Eq. (\ref{88}), what is explained in the next section.

\begin{figure}[h]%
\begin{center}\includegraphics[width=0.9\linewidth]{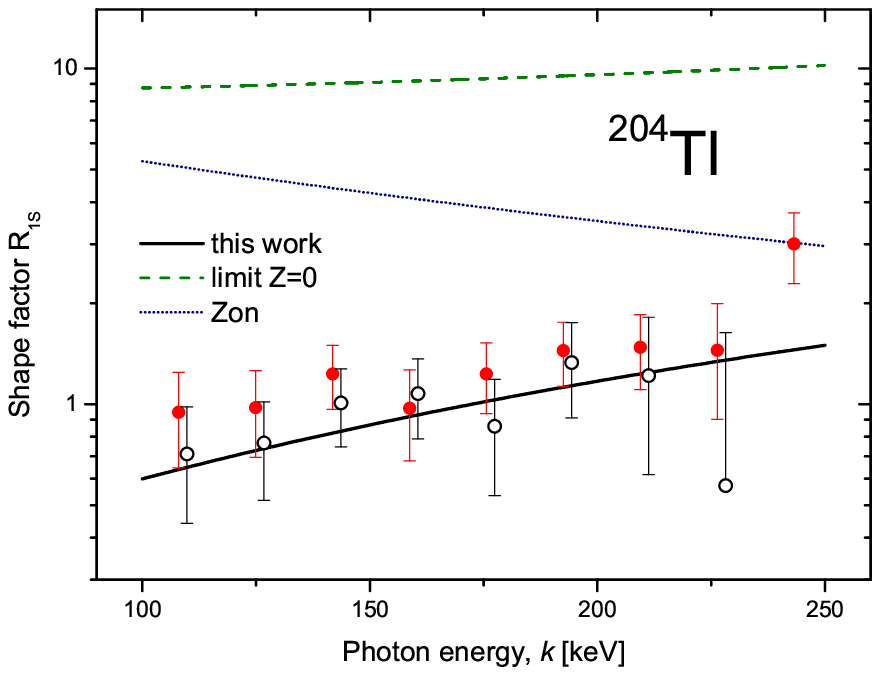}\end{center}%
\caption{\label{fig2}
The shape factor for the 1S component of the first forbidden unique REC in $^{204}$Tl.
The points with error bars represent the experimental values form Ref.\cite{kurcewicz}.
The meaning of theoretical lines is the same as in Fig.1. The value of $Q$ is
$264.0$ keV.}
\end{figure}

The case of $^{204}$Tl ($Q_{EC} = 346.5$ keV, $T_{1/2} = 3.78$ years) is illustrated
in Fig.~2. Again, our results  perfectly agree with experimental data, while the
results of Zon and Rapoport overestimate the spectrum by a factor
of about 4 \cite{kurcewicz}.  As expected, the influence of Coulomb field is much
stronger in case of thallium ($Z = 81$) than in case of calcium ($Z = 20$).
The nonrelativistic limit would overestimate the intensity of the REC spectrum
by a factor of about 10.

\section{Comparison to previous calculations}
We have verified numerically that Eq. (\ref{84}) agrees with former results obtained
in Ref. \cite{glauber, martin, zrap, zon} for the allowed and first forbidden
non unique EC transitions. However, in contrast to
previous works \cite{zrap, zon, kalinowski}, our results correctly describe the
known experimental data for the first forbidden unique decays.
The difference, in our opinion, is due to incorrect approximations adopted in the
Refs.\cite{zrap, zon}. If we use the Coulomb gauge for the emitted photon,
Eqs. (\ref{35p},\ref{36p}), then the approximation of Eq. (\ref{65})
for the Coulomb Green's function is not valid, because the integral
with the neglected remainder, which is proportional to $\Theta(r-r')$ is as important
as the integral with the approximated form in  Eq. (\ref{65}). As an consequence 
of this approximation in the Coulomb gauge, ${\cal R}_{E1}(S_{1/2}\rightarrow P_{3/2})$
in Eq. (\ref{75}) would contain $j_0(k\,r)$,  and the integral would diverge
for small values of $r$, as the integral in the function $B_{21}$ of Ref. \cite{zon} does.
With the length gauge, Eqs. (\ref{35},\ref{36}), where $E1$ photon wave
function contains a combination of $j_1(k\,r)$ and $j_2(k\,r)$ functions,
the $r$-integral in Eq. (\ref{75}) is finite, as it should. Moreover, the probability
amplitude for the radiation from the nucleus is strongly suppressed,
because the photon wave function is much smaller within the nucleus.
However, we can not exclude completely the nuclear contribution
if there are close-lying excited states in the daughter or in the parent nucleus.
In such a case, the nuclear radiation should be obtained using the same gauge.
Since the photon wavelength is much larger than the size of the nucleus,
the dipole approximation is allowed. Then, the coupling of nucleus to
the electromagnetic field in the length gauge
takes the form $-\vec d\cdot \vec E - \vec\mu\cdot\vec B$,
and this form should be used for the calculation of
the photon radiation from the nucleus.

\section{Summary}
We have recalculated the rate of radiative orbital electron capture by the
atomic nucleus. We applied the length gauge for the radiated photon,
because it suppresses the nuclear radiation and
substantially simplifies the calculations.
By the use of a convenient form of the Dirac-Coulomb Green function
from Ref. \cite{dhbk}, the results for
the capture of the 1S electron can easily be generalized
for the arbitrary state of the captured electron.
The results obtained are found to be in good agreement with experimental data
for allowed, first forbidden and first forbidden unique transitions,
in particular with the results for $^{41}$Ca and $^{204}$Tl.

There is another case of the first forbidden unique EC transition,
the decay of $^{81}$Kr, which can be used to verify predictions
presented in this paper. An experiment with a sample of
$^{81}$Kr is being undertaken, and its outcome will be published soon.

\section*{Acknowledgments}
We wish to acknowledge fruitful discussions with J. \.Zylicz, P.
Chankowski and  P.~Hornsh\o j.

\end{document}